\documentclass[aps,prx,twocolumn,amsmath,amssymb,nofootinbib,superscriptaddress,floatfix,reprint,longbibliography]{revtex4-2}
\usepackage[dvips]{graphicx}
\usepackage{latexsym}
\usepackage{amsmath}
\usepackage{amsfonts}
\usepackage{amssymb}
\usepackage{bm}
\usepackage{color}
\usepackage{txfonts}
\usepackage{float}
\usepackage{url}
\usepackage[colorlinks=true, urlcolor=blue, linkcolor=blue, citecolor=blue]{hyperref}
\usepackage{ulem}
\usepackage{mhchem}
\usepackage{physics}
\begin{document}
	\newcommand{\fig}[2]{\includegraphics[width=#1]{#2}}
	\newcommand{\la}{{\langle}}
	\newcommand{\ra}{{\rangle}}
	\newcommand{\dg}{{\dagger}}
	\newcommand{\upa}{{\uparrow}}
	\newcommand{\dna}{{\downarrow}}
	\newcommand{\ab}{{\alpha\beta}}
	\newcommand{\ias}{{i\alpha\sigma}}
	\newcommand{\ibs}{{i\beta\sigma}}
	\newcommand{\hH}{\hat{H}}
	\newcommand{\hn}{\hat{n}}
	\newcommand{\hc}{{\hat{\chi}}}
	\newcommand{\hU}{{\hat{U}}}
	\newcommand{\hV}{{\hat{V}}}
	\newcommand{\br}{{\bf r}}
	\newcommand{\bk}{{{\bf k}}}
	\newcommand{\bq}{{{\bf q}}}
	\def\gsim{~\rlap{$>$}{\lower 1.0ex\hbox{$\sim$}}}
	\setlength{\unitlength}{1mm}
	\newcommand{{\vhf}}{$\chi^\text{v}_f$}
	\newcommand{{\vhd}}{$\chi^\text{v}_d$}
	\newcommand{{\vpd}}{$\Delta^\text{v}_d$}
	\newcommand{{\ved}}{$\epsilon^\text{v}_d$}
	\newcommand{{\vved}}{$\varepsilon^\text{v}_d$}
	\newcommand{\pprl}{Phys. Rev. Lett. \ }
	\newcommand{\pprb}{Phys. Rev. {B}}

\title {General Theory of  Josephson  Diodes}
\author{Yi Zhang}
\affiliation{Department of Physics, Shanghai University, Shanghai 200444, China}
\affiliation{Kavli Institute of Theoretical Sciences, University of Chinese Academy of Sciences,
	Beijing, 100190, China}

\author{Yuhao Gu}
\affiliation{Beijing National Laboratory for Condensed Matter Physics and Institute of Physics,
	Chinese Academy of Sciences, Beijing 100190, China}

\author{Pengfei Li}
\affiliation{Beijing National Laboratory for Condensed Matter Physics and Institute of Physics,
	Chinese Academy of Sciences, Beijing 100190, China}

\author{Jiangping Hu}
\email{jphu@iphy.ac.cn}
\affiliation{Beijing National Laboratory for Condensed Matter Physics and Institute of Physics,
	Chinese Academy of Sciences, Beijing 100190, China}
\affiliation{Kavli Institute of Theoretical Sciences, University of Chinese Academy of Sciences,
	Beijing, 100190, China}

\author{Kun Jiang}
\email{jiangkun@iphy.ac.cn}
\affiliation{Beijing National Laboratory for Condensed Matter Physics and Institute of Physics,
	Chinese Academy of Sciences, Beijing 100190, China}

\date{\today}

\begin{abstract}
Motivated by recent progress in the superconductivity nonreciprocal phenomena, we study the general theory of  Josephson diodes. The central ingredient for Josephson diodes is the asymmetric proximity process inside the tunneling barrier.
From the symmetry breaking point of view, there are two types of Josephson  diodes, inversion breaking and time-reversal breaking. For the inversion breaking  case, applying voltage bias could effectively tune the proximity process like the voltage-dependent Rashba coupling or electric polarization giving rise to $I_c(V)\neq I_c(-V)$ and $I_{r+}\neq I_{r-}$.
For the time-reversal breaking case, the current flow could adjust the internal time-reversal breaking field like magnetism or time-reversal breaking electron-electron pairing, which leads to $I_{c+}\neq I_{c-}$. All these results provide a complete understanding and the general principles of realizing Josephson diodes, especially the recently found NbSe$_2$/Nb$_3$Br$_8$/NbSe$_2$ Josephson diodes.
\end{abstract}
\maketitle

\section{Introduction}
As a macroscopic quantum phenomenon, superconductivity is one of  the most important subjects in condensed matter physics \cite{schrieffer,tinkham,degennes}. The central ingredients for a superconductor (SC) are the electron-electron pairing and phase coherence, which gives rise to the absence of resistivity and the Meissner effect. 
Brian Josephson elegantly linked the pairing condensation and phase coherence with the supercurrent generation between two weak-linked superconductors, which is now known as the Josephson effect or Josephson junction (JJ) \cite{andeson,josephson}. The emergence of the Josephson effect enables the wide applications of superconductivity, like the superconducting quantum interference devices (SQUIDs), frequency detectors etc \cite{tinkham}. 

However, compared with modern semiconductor electronic devices, the devices based on superconducting current are still very limited.  For the normal electric current, a semiconductor  p-n junction,  known as the diode, conducts current primarily in one direction. This non-reciprocal charge transport has multiple usages including rectification of current, detection of radio signals, temperature sensor etc. It also serves as the basic component of computer memory and logic circuit, which is essential for computer development.
All these make the diode become one of the key devices in the semiconductor industry~\cite{sze}.  Thus, a natural question for Josephson junction arises :  is there a diode for the superconducting current?  We will name such a diode as  Josephson diode (JD). 

Most recently,  a Josephson diode without a magnetic field  has been  observed in an inversion asymmetric NbSe$_2$/Nb$_3$Br$_8$/NbSe$_2$ (NSB) heterostructure~\cite{ali}, which experimentally shows the critical current in the positive direction deviates from the negative one in a JJ for the first time. Besides the JJ, the bulk superconductor using Nb/V/Ta superlattice has also been found to have a similar diode effect under magnetic field~\cite{ando} and many new systems have been reported to be non-reciprocal in both JJs and bulk superconductors~\cite{NiTe2,tTLG,tbg,kim_2021,Baumgartner_2022,sde_NbSe2,Bau_2022}. All these findings not only enrich the zoo of superconductivity phenomena, but also point to a new direction in superconducting electronics like superconducting computer chips, direction-selective quantum sensors, rectifier and other quantum devices \cite{ali,ando}. Last but not least, as we will show in this work, one class of the diode effect is closely related to the time-reversal symmetry, therefore, it can potentially provide a new method to detect the time-reversal symmetry breaking in the superconducting system.
Since constructing a diode using bulk SC requires the superconducting disfavored time-reversal symmetry breaking as discussed below, we will focus on the Josephson diode effect by engineering its more flexible barrier part in this work.

Historically, the first theoretical proposal for Josephson diodes stems from the SC analogy of p-n junctions by the electron and hole doped SCs close to an SC-Mott-insulator transition \cite{hu}.  
In addition, the anomalous Josephson effect closely related to the so-called $\phi_0$ Josephson state has been studied intensively~\cite{buzdin_2008,yuli_2013,yuli_2014,julia_2015,kou_2016,Marco_2019,phi0_2013,phi0_2016,phi0_2017,phi0_2018,phi0_2020,phi0_2021,phi0_2022}, which is one possible mechanism to realize the nonreciprocal transporting effect in the JJs.
Recently, the nonreciprocal Josephson effect  utilizing the  charging asymmetry effect has been studied by semiclassical approaches \cite{nagaosa1}.  Using the magnetochiral anisotropy, the non-reciprocal responses and superconducting diode effects under the external magnetic field have been investigated both theoretically and experimentally \cite{ando,hejun,nagaosa17,nagaosa18,yanse,yuan,fu_2022,tTLGt}. Similarly, the asymmetric Fermi velocities of topological material edge states under external magnetic fields have also been proposed to have a nonreciprocal effect \cite{law}. 
The experiment on NSB~\cite{ali} goes beyond the above theoretical considerations. Thus, it calls for a  broader theory for the Josephson diode.

\begin{figure}
	\begin{center}
		\fig{3.4in}{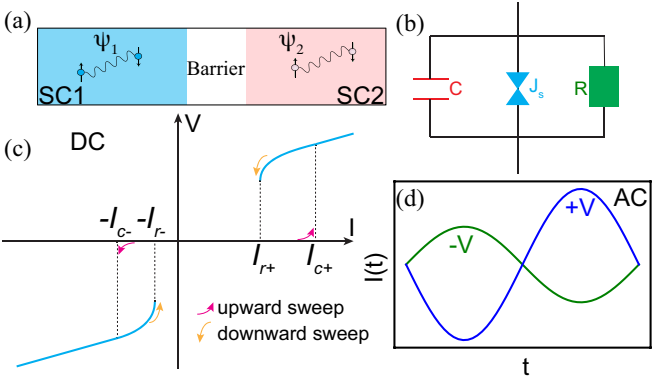}\caption{(a) A Josephson junction is constructed by two SCs $\psi_{1/2}$ sandwiched with a tunneling barrier.
		 (b) The physical Josephson junction can be modeled by an ideal Josephson junction $J_s$ shunted with a resistance and a capacitance. The capacitance leads to the hysteresis curve in I-V characteristic. 
		(c) General I-V curves for a Josephson diode in a DC measurement. During the upward current sweep, there are two critical currents $I_{c+}$ (positive direction)  and $-I_{c-}$ (negative direction) from the SC regime to the dissipative regime,  where $I_{c+}\neq I_{c-}$ shows a diode effect.  During the downward current sweep, there are also two return critical currents $I_{r+}$ (positive direction) and $-I_{r-}$ (negative direction) from the dissipative regime to SC regime, where  $I_{r+}\neq I_{r-}$ shows another diode effect. (d), The diode effect in a AC  Josephson junction, with $I_c(V)\ne I_c(-V)$. 
			\label{definition}}
	\end{center}
	\vskip-0.5cm
\end{figure}

\section{Josephson Diode definition and types} 
Generally speaking, a Josephson junction is constructed by two SCs sandwiched with a non-SC tunneling barrier, as illustrated in Fig.\ref{definition}(a). 
Phenomenologically, the Josephson relation can be understood from the Ginzburg-Landau (G-L) theory \cite{ketterson,degennes,tinkham}, which is described by two macroscopic pairing potentials $\psi_1$ and $\psi_2$ for two SCs \cite{ketterson}. The G-L boundary condition at the interface can be written as 
\begin{eqnarray}
	\frac{\partial \psi_1}{\partial z}&=&\frac{\psi_2}{b} \\
	\frac{\partial \psi_2}{\partial z}&=&-\frac{\psi_1}{b} 
\end{eqnarray}
where the length $b$ is a phenomenological length describing the tunneling barrier.
Then the Josephson current can be found from G-L equations as
\begin{eqnarray}
	I=\frac{2e \hbar}{m^* b} |\psi_1||\psi_2| \sin(\phi_1-\phi_2)
	\label{DC}
\end{eqnarray}
where the $\phi_{1/2}$ are the corresponding phases for $\psi_{1/2}$ respectively. $m^*$ is the effective mass for SCs. 
This is the DC Josephson effect. From this equation, we can easily conclude that the critical current $I_c$ depends on length $b$ and the amplitudes $|\psi_{1/2}|$.
Besides the DC Josephson effect, Josephson predicted another AC Josephson effect, where the time-dependence of phase difference is related to the voltage bias
by $\frac{d(\phi_1-\phi_2)}{dt}=\frac{2eV}{\hbar}$. This difference leads to AC Josephson effect
\begin{eqnarray}
    I(t)=I_c \sin(\frac{2e}{\hbar}Vt+\Delta \phi_0)
    \label{AC}
\end{eqnarray}
where $\Delta \phi_0$ is the phase difference at zero voltage.

Because of the sandwich structure, a realistic Josephson junction consists of an ideal JJ described by Eq.\ref{AC}, an effective resistance and a capacitance connected in parallel as shown in Fig.~\ref{definition}(b), namely the RCSJ model \cite{tinkham,ketterson}. Hence, the realistic Josephson junction shows a more complicated I-V characteristic like the hysteresis curves in Fig.~\ref{definition}(c). 
During the upward current sweep, after reaching the critical current $I_c$, the Josephson junction loses its non-dissipative SC property without voltage and resistance and enters a dissipative regime with finite voltage and resistance. On the other hand, during the downward current sweep, the capacitance has been charged at finite voltage stage. This charged capacitance leads to another critical current namely the return current $I_r$. The difference between $I_r$ and $I_c$ gives rise to the hysteresis behavior of the  I-V characteristic curve in Fig.~\ref{definition}(c).

 In analogy to the voltage controlled p-n junction, if the critical current in the positive direction $I_{c+}$ deviates from the critical current in the negative direction $I_{c-}$, a Josephson diode effect is achieved with $I_{c+}\neq I_{c-}$, as illustrated in Fig.\ref{definition}(c). 
 Similarly, if  the $I_{r+}$ in the positive direction is different from the $I_{r-}$ in the negative direction, another Josephson diode effect with $I_{r+}\neq I_{r-}$ emerges. And this $I_{r+}\neq I_{r-}$ is closely related to the finite voltage history during the downward sweep.  Both $I_{c+}\neq I_{c-}$  and $I_{r+}\neq I_{r-}$ have been observed in recent Josephson diode experiments \cite{ali,NiTe2}. As an extension of the voltage dependent $I_{r+}\neq I_{r-}$ effect, the AC Josephson junction controlled by voltage bias can also show a diode effect. As illustrated in Fig.\ref{definition}(c), if the $I_c$ depends on the voltage with $I_c(+V)\neq I_c(-V)$, another Josephson diode controlled by voltage is achieved, which is similar to the proposal in Ref.~\onlinecite{hu}.

From the symmetry point of view, the key symmetry of any directional dependent diode effect is the inversion symmetry ${\cal I}$. For instance, 
the built-in potential of p-n junction induced by ${\cal I}$ symmetry breaking leads to the competition with the external voltage.
Similarly, the voltage dependent $I_{r+}\neq I_{r-}$ and $I_c(+V)\neq I_c(-V)$ diode effects must require a ${\cal I}$ symmetry breaking.
However, for the superconducting diode effect at the zero voltage, the ${\cal I}$ breaking is only the minimal requirement. 
The time-reversal symmetry ${\cal T}$ for Josephson diode is another crucial symmetry. Owing to Onsager reciprocal relations, the responses of a time-reversal invariant system under two opposite external fields are related to each other by the ${\cal T}$ operation.
As the ${\cal T}$ broken current $I=\frac{dq}{dt}$ is the only external field for the Josephson junction upward sweep, the Hamiltonian only depends on current $\hat{H}(I)$. We can first assume $\hat{H}(I=0)$ is ${\cal T}$ invariant. Then, the Hamiltonian at positive current $\hat{H}(+I)$ is related to the negative one  $\hat{H}(-I)$ by a  ${\cal T}$ operation, which ensures $I_{c+}=I_{c-}$. The $I_{c+}\neq I_{c-}$ phenomenon in JD directly breaks the symmetry relation indicating the ${\cal T}$ breaking for Josephson diode at zero current.


\begin{figure}
	\begin{center}
		\fig{3.4in}{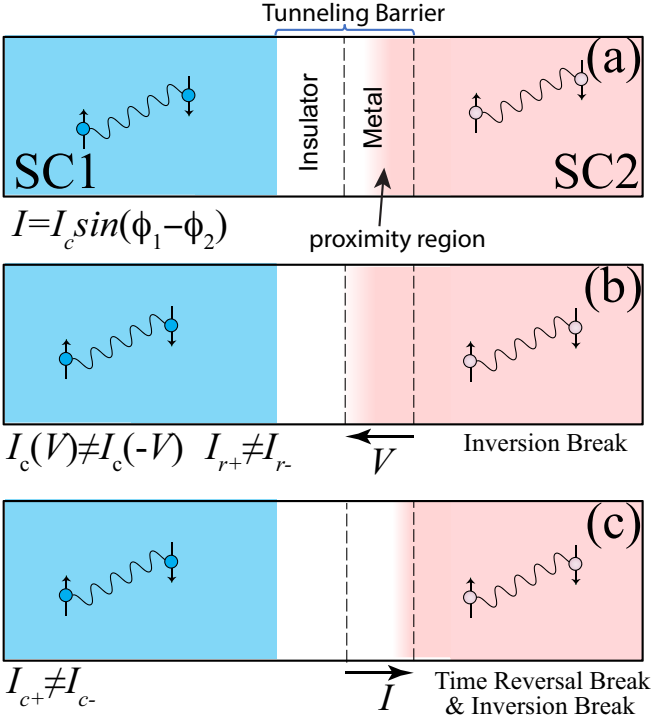}\caption{(a) The geometric setup for a Josephson diode in an extreme limit, which is formed by SC $\mathbf{\Delta}_{1}$  on the left, SC $\mathbf{\Delta}_{2}$  on the right and the tunneling barrier (TB). The  Josephson current is determined by $I=I_c \sin(\phi_1-\phi_2)$, where $\phi_{1/2}$ is the phase and $I_c$ is the critical current. The tunneling barrier layer in the  extreme limit is formed by an insulator layer and a metallic layer (N).  And the $\mathbf{\Delta}_{1}$ influences the N layer by forming the proximity region.
			(b) An inversion breaking JD is controlled by voltage with $I_{c}(V)\neq I_{c}(-V)$ or $I_{r+}\neq I_{r-}$, which effectively adjusts the proximity region. To illustrate this point, the voltage (positive) enlarges the proximity region, which increases the $I_{c}$.
			(c) A time reversal breaking JD (with inversion breaking) controlled by current flow I with $I_{c+}\neq I_{c-}$. To illustrate this point, the current follow in the positive direction reduces the proximity region, which decreases the $I_{c}$.
			\label{fig1}}
	\end{center}
	\vskip-0.5cm
\end{figure}

Based on the above discussion, we can find that the most convenient way to achieve Josephson diode is through designing the barrier part with proper symmetry breaking. Since we normally use common SCs in JJ constructions, engineering the barrier is equivalent to changing the phenomenological length $b$ effectively, which leads to the change in the critical current from Eq.\ref{DC}.
In order to achieve this goal, we propose the Josephson diode design in Fig.\ref{fig1}.  As shown in Fig.\ref{fig1}(a), a Josephson  diode is formed by a tunneling barrier (TB) and two SCs on the left ($\mathbf{\Delta}_{1}$) and  the right ($\mathbf{\Delta}_{2}$) respectively. Since the minimal symmetry requirement is the ${\cal I}$ symmetry breaking. Therefore, the coupling between TB and $\mathbf{\Delta}_{1}$ must be different from the coupling between TB and $\mathbf{\Delta}_{2}$. To simplify our discussion, we will take an extreme limit, where the TB layer is formed by an insulator layer  and a metallic layer (N layer), as illustrated in Fig.\ref{fig1}(a). And the Nb$_3$Br$_8$ barrier in Ref. \cite{ali} NSB heterostructure indeed belongs to this case, which will be discussed below \cite{sm}. Owing to the metallic nature, the $\mathbf{\Delta}_{2}$ will induce superconducting pairing into the N layer by generating an SC proximity region as illustrated in Fig.\ref{fig1}(a). This SC proximity region can serve as the effective ``depletion" region as in the semiconductor p-n junction. Clearly, tuning the proximity region is equivalent to tuning the effective length of TB and the Josephson coupling between $\mathbf{\Delta}_{1}$ and $\mathbf{\Delta}_{2}$. Hence, if we can control this proximity region, a Josephson diode can be easily realized as illustrated in Fig.\ref{fig1} (b),(c). For example, if the current $I$ reduces the proximity region in the positive direction while enlarging this region in the negative direction, an $I_{c+}<I_{c-}$ effect is achieved as illustrated in Fig.\ref{fig1}(c).
In short, there are two types of Josephson diodes, inversion breaking JD with $I_{r+}\neq I_{r-}$ or $I_c(+V)\neq I_c(-V)$ and time reversal breaking JD with $I_{c+}\neq I_{c-}$  from the symmetry point of view. We will discuss them separately in the following sections.

\section{Inversion breaking JD}
We start from the physics of inversion symmetry breaking Josephson diode.
For an ${\cal I}$ breaking JD, the essential part is to find voltage-dependent quantities in the TB.  In the conventional p-n junctions, the depletion region is formed by diffusion between electrons from n-doped region and holes from the p-doped region. Then, the built-in potential between holes and electrons inside the depletion region competes with the external voltage giving rise to the nonreciprocal transport. The Josephson diode using the hole and electron-doped SC is also based on the similar built-in potential by electrons and holes, where the depletion region is formed by a self-organized Mott insulator region \cite{hu}. 
This JD belongs to the ${\cal I}$ breaking  Josephson diode \cite{hu}. 

Additionally, there are many other quantities that can be controlled by voltage, for example the Rashba spin orbital coupling (SOC) $\alpha  \pmb{\sigma} \times \mathbf{p}$ \cite{rashba_59,rashba_84,rashba_review,vasko}.
The Rashba SOC results from the ${\cal I}$ symmetry breaking induced interfacial electric field $E$ at material interfaces or two-dimensional metallic planes. The external voltage can adjust the asymmetric crystal potential, which  tunes the electric field $E \propto -\nabla V$.
Therefore, applying a voltage can efficiently change the magnitude of $\alpha$ \cite{rashba_review,quantum_well,sto,rashba_tune,rashba_tune2}. Voltage-controlled Rashba effect has been realized in many semiconductor heterostructures such as the quantum well consisting of single HgTe~\cite{rashba_tune}, single InAs~\cite{rashba_tune2}, inverted InAlAs/InGaAs heterostructure~\cite{quantum_well} and the interface of SrTiO$_3$/LaAlO$_3$~\cite{sto}. Taking the InAlAs/InGaAs heterostructure as an example, the Rashba constant $\alpha$ can be efficiently tuned in the range of about (0.64$\times$10$^{-11}$eV m, 0.93$\times$10$^{-11}$eV m)~\cite{quantum_well}.
And the magnitude of $\alpha$ will influence the proximity region owing to the changing of Fermi momentum $k_F$ and the spin texture along the Fermi surfaces (FSs).

\begin{figure}
	\begin{center}
		\fig{3.4in}{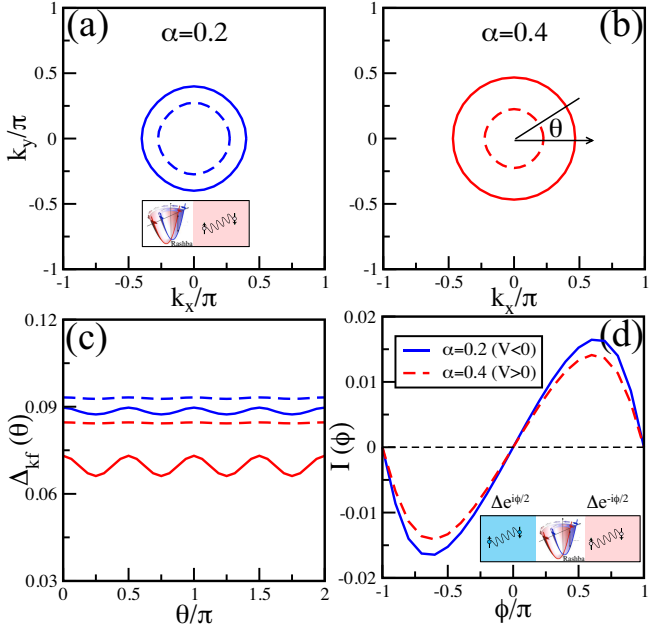}\caption{(a) Fermi surface for TB with $\alpha=0.2$. The inset illustrates the proximity process leading to the pairing amplitude in (c). (b) Fermi surface for TB with $\alpha=0.4$. (c) Spin singlet pairing strengths along each FS for $\alpha=0.2, 0.4$. $\theta$ is the angle along each FS as defined in (a), (b). (d) The Josephson currents $I(\phi)$ in unit of $\frac{2e}{\hbar}$ for the Rashba JD for $\alpha=0.2$ and $\alpha=0.4$ and the inset shows the setup for calculating the current. The other parameters are set as $t_R=t_L=1.0$, $\Delta_1=\Delta_2=0.2$, $t_{RB}=0.8$, $t_{LB}=0.6$, $\mu=-3.0$. In the calculation, we set the thickness of SC on both sides to be 20.
			\label{fig2}}
	\end{center}
	\vskip-0.5cm
\end{figure}

To justify this theory, we start from the proximity process between the right SC $\mathbf{\Delta_{2}}$ and TB with Rashba SOC, as illustrated in Fig.2(a). 
The Hamiltonian can be written as  $H_{0}=H_{R}+H_{TB}+H_{RTB}$. The $H_{R}$ describes the right SC with a cubic lattice and the s-wave pairing for the spinor $c_i=(c_{i,\uparrow},c_{i,\downarrow})^T$  as
\begin{eqnarray}
 	H_{R}=-t_{R}\sum_{<ij>}c^{\dagger}_ic_j+\mathbf{\Delta}_{2}\sum_i c_{i\uparrow}c_{i\downarrow}+h.c.
\end{eqnarray}
The $H_{TB}$ describes the TB layer with a square lattice and Rashba SOC for the spinor $f_i=(f_{i,\uparrow},f_{i,\downarrow})^T$ as
\begin{eqnarray}
	H_{TB}=-t_{TB}\sum_{<ij>}f^{\dagger}_if_j-i\alpha\sum_{<ij>}f^{\dagger}_i(\pmb{\sigma}\times\mathbf{d}_{ij})_zf_j+h.c.
\end{eqnarray}
where the $\mathbf{d}_{ij}$ is the unit vector from site i to site j. The coupling between them is described by $H_{RTB}$ as
\begin{eqnarray}
	H_{RTB}=-t_{RB}\sum_{<ij>}f^{\dagger}_ic_j+h.c.
\end{eqnarray}
In this setup, the current is flowing in the $z$ direction as well as the voltage drop.
We assume that $\alpha$ relates to voltage bias $V$ by a phenomenological coupling $c_{\alpha}$ with $\alpha=\alpha_0+c_{\alpha}V$, where $\alpha_0$ is Rashba constant without voltage. If we further assume $\alpha_0$=0.3 and choose proper $c_\alpha$ and voltage $V_0$, we can have $\alpha(-V_0)=0.2$ and $\alpha(V_0)=0.4$, as we used in the calculation in Fig.~\ref{fig2}.
Owing to Rashba SOC, the spin-degenerate FSs split into two helical FSs with spin-momentum locking as $\pmb{\sigma} \times \mathbf{k}$. 
The  $\delta k_F$ difference between two split FSs depends on the magnitude of $\alpha$. By comparing the FSs at Fig.\ref{fig2}(a),(b), we can find $\delta k_F$  for $\alpha=0.2$ is smaller than $\alpha=0.4$. 
Then, we can compare the effective pairing strength $<f_{k\uparrow}^\dagger f_{-k\downarrow}^\dagger>$ for each proximity process. The effective pairing $\Delta(\theta)$s along the TB FSs in Fig.\ref{fig2}(c) show
that  $\alpha=0.2$  obtains a much larger pairing than $\alpha=0.4$. 
As we know, the phase coherent length of the Cooper pairs leaking from the SC to the metal due to the Andreev reflection of the electron with energy $\epsilon<\Delta$ is given by $L_c=\min(\sqrt{\hbar D/\epsilon},L_{\phi})$~\cite{andreev_pro}, with D the diffusion constant of the metal phase, and $L_{\phi}$ the single electron phase coherent length usually determined by the disorder. Therefore, when $\epsilon\ll\Delta,\hbar D$ so that $L_c$ is larger than the size of the TB, corresponding to the short junction case as we consider here, the proximity process between metal and SC is coming from the Andreev reflection~\cite{andreev,andreev_pro}.
The mismatching between the metal $k_F$ and SC momentum gives rise to this  pairing strength difference between different $\alpha$. Since the $k_F$ of $\alpha=0.2$ is much closer to the $k_F$ of $\mathbf{\Delta_{2}}$, a much larger pairing is obtained.
Hence, adjusting Rashba SOC can efficiently adjust the proximity region.

To calculate the Josephson  effect, we still need to couple $H_0$ with the left part $\mathbf{\Delta_{1}}$. The Hamiltonian $H_{L}$ is similar to $H_{R}$ as
\begin{eqnarray}
	H_{L}=-t_{L}\sum_{<ij>}c^{\dagger}_ic_j+\mathbf{\Delta}_{1}\sum_i c_{i\uparrow}c_{i\downarrow}+h.c.
\end{eqnarray}
and the coupling with TB is written as 
\begin{eqnarray}
	H_{LTB}=-t_{LB}\sum_{<ij>}f^{\dagger}_ic_j+h.c.
\end{eqnarray}
The inversion symmetry breaking can be simulated by $t_{LB}\neq t_{RB}$. It seems that this setup ignores the insulator layer. However, setting $t_{LB}\neq t_{RB}$  is just equivalent to integrating out the insulator layer degree of freedom.
We can further introduce a phase  into the SCs as $\mathbf{\Delta}_{1}=\Delta_0 e^{i\phi/2}$ and $\mathbf{\Delta}_{2}=\Delta_0 e^{-i\phi/2}$. Then the supercurrent through the junction is related to the total Hamiltonian $H_{t}=H_{0}+H_{L}+H_{LB}$ by
\begin{eqnarray}
	I(\phi)=\frac{2e}{\hbar}\partial_\phi \sum_{n}f(\epsilon_n) \epsilon_n(\phi)
	\label{current}
\end{eqnarray}
where the $\epsilon_n$ is the n-th eigenvalue for $H_t$ at the phase $\phi$ \cite{beenakker}. 
The $I(\phi)$ is calculated as the function of $\phi$ in Fig.\ref{fig2}(d).  The results in Fig.\ref{fig2}(d) demonstrate the critical current $I_c$ for Rashba Josephson junction decreases with increasing $\alpha$.
Therefore, a  voltage controlled JD with $I_{c}(V)\neq I_{c}(-V)$ can be realized by the voltage dependent Rashba SOC, which can be detected through the AC Josephson measurement as discussed above. The external voltage competes with the internal interfacial voltage at the interfaces, leading to a tunable Rashba SOC. This Rashba SOC dependent critical current has also been discussed using Green’s function method~\cite{jc_soc}.

\begin{figure}
	\begin{center}
		\fig{3.4in}{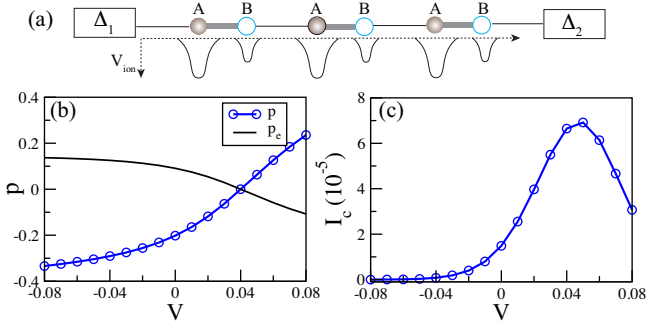}\caption{(a) A schematic diagram for a Rice-Mele model JD, where the tunneling barrier is a 1D Rice-Mele chain. The Rice-Mele chain is formed by sublattices A and B with different on-site potential $V_{ion}$. The strong bonds (thick gray lines)  and weak bonds (thin black lines) are linked to A,B alternatively as in Su-Shrieffer-Heeger model.
			(b) The polarization density $p$ (unit of $e$) as a function of voltage with the electron part contribution $p_e$ for the bulk Rice-Mele model. (c) The critical current $I_c$ (unit of $\frac{2e}{\hbar}$) for Rice-Mele chain as a function of voltage, where the $I_c(+V)\neq I_c(-V)$ demonstrates the diode effect. The other parameters are set as $t_R=t_L=t=1.0$, $\Delta_1=\Delta_2=0.2$, $t_{RB}=t_{LB}=0.8$, $\delta t_{0}=0.2$, $\mu=-1.0$, $Q=0.5$, $\beta=5.0$. 
			\label{figr_m}}
	\end{center}
	\vskip-0.5cm
\end{figure}

Besides the voltage-dependent Rashba coupling, another common voltage-controlled phenomenon is electric polarization $p$ in ferroelectricity. Inside the ferroelectric materials, their electric polarization $p$ highly depends on the external voltage, which can be used for the inversion broken JD. The simplest model for ferroelectricity is the Rice-Mele model~\cite{rice_mele,vander_1993,nagaosa_2004,Resta_1993,Resta_1994,niu_2010}. As illustrated in Fig.~\ref{figr_m}, we construct the JJ using the Rice-Mele chain as the tunneling barrier.
The Rice-Mele model is the extension of Su-Shrieffer-Heeger (SSH) model with different sub-lattice potential $V_{ion}(i)$ for each site $i$~\cite{ssh,niu_2010}.
\begin{eqnarray}
	H_{RM}=\sum_i \left(\frac{t}{2}+(-1)^i \frac{\delta t}{2}\right)\left(f_{i}^\dagger f_{i+1}+h.c.\right)+V_{ion}(i) f_{i}^\dagger f_{i}
\end{eqnarray}
The $t+\delta t$ and $t-\delta t$ describe the alternating strong bond and weak bond along the chain, as in the SSH model. The $V_{ion}(i)$ is the on-site sub-lattice potential.
We set the $V_{ion}(A)=Q,V_{ion}(B)=-Q$, where $Q$ describes the on-site potential difference between A and B. 
When $\delta t=0$, the Rice-Mele model still has the inversion symmetry with respect to A or B. Hence, the Rice-Mele chain breaks the ${\cal I}$ symmetry with polarization $p$ only when both $\delta t$ and $Q$ are finite. 
To describe the ferroelectricity of this model, we also need to introduce one phenomenological parameter $\beta$ describing the polarizability of the Rice-Mele chain and the alternating hopping difference $\delta t$ is related to $V$ by $\delta t_0-\beta V$, with initial bond difference $\delta t_0$.  

Since the polarization density $p$ at $\delta t=0$ is exactly zero, we take this point as the reference point. Then, $p$ is calculated as a function of voltage as shown in Fig.\ref{figr_m}(b). 
From the modern theory of polarization,  the polarization has two parts $p=p_{ion}+p_{e}$, where $p_{ion}$ is from ion dipole moment and $p_{e}$ is from the electron part and the $p_{e}$ can be calculated using the Wannier center and Berry phase method~\cite{Resta_1994,niu_2010} while
the ion part is from ion charge and bond lengths described in the supplementary material~\cite{sm}. 
From Fig.\ref{figr_m}(b), we can find that external voltage changes the $p$ continuously as in ferroelectic materials.
Because of the spontaneous electric polarization $p_0$, a finite $p$ is found when $V=0$.
Then, the critical currents for each configuration are calculated using Eq.\ref{current}.
As shown in Fig.\ref{figr_m}(c), the critical current at negative voltage almost vanishes after $V=-0.04$ while the $I_c$ is still finite at positive V.
Therefore, using the Rice-Mele chain, a Josephson diode with $I_c(+V)\neq I_c(-V)$ and AC Josephson diode can be achieved.

In a short summary, through engineering the tunneling barrier by Rashba SOC and electric polarization, we demonstrate that the critical current of JJ can be adjusted by voltage. This scenario can be used to construct JD with $I_c(+V)\neq I_c(-V)$ and  $I_{r+}\neq I_{r-}$, which is also related to recent experimental findings as discussed below.

\section{Time reversal breaking JD}
Besides the above inversion breaking JD, there is another type of JD by further breaking ${\cal T}$ symmetry.  To achieve this goal, the TB layer must break ${\cal T}$ owing to Onsager reciprocal relations and symmetry requirement discussed in the introduction section.
The simplest ${\cal T}$ breaking phenomenon in solid state physics is magnetism. 
Hence, assuming the TB layer contains the internal magnetism, if the proximity region can be adjusted by tuning the magnetization, a  ${\cal T}$ breaking JD can be achieved.

To simulate this magnetic order, we can add an $s$-$d$ exchange coupling term $\sum_i f^{\dagger}_i M\cdot\pmb{\sigma} f_i$ \cite{double-exchange,nagaosa_prb} into the barrier microscopic Hamiltonian $H_{TB}$, where $\textbf{M}$ describes the localized spin and $\pmb{\sigma}$ describes the spin of conduction electrons. As an extension to current induced $I_{c+}\neq I_{c-}$, we first use magnetism amplitude under magnetic field $B$ as an example to demonstrate the tunability of ${\cal T}$ broken JJ, which can be viewed as an extension of JD. In this case, we assume $M=(M_0+c_MB)\textbf{e}_z$, where $M_0$ is the initial magnetic value and $c_M$ is effective susceptibility in response to the magnetic field.

\begin{figure}
	\begin{center}
		\fig{3.4in}{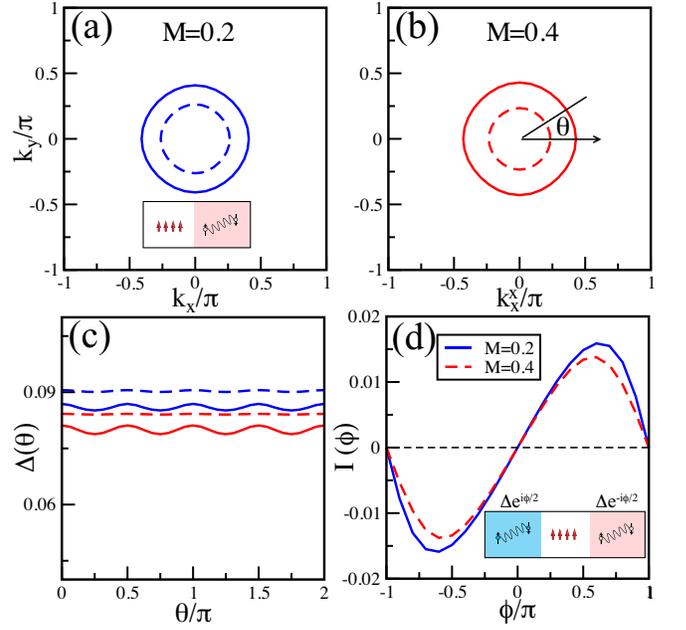}\caption{(a) Fermi surface for TB with $M=0.2$. The inset illustrates the proximity process leading to the pairing amplitude in (c). (b) Fermi surface for TB with $M=0.4$. (c) Spin singlet pairing strengths along each FS for $M$=0.2, 0.4. $\theta$ is the angle along each FS as defined in (a), (b). (d) The Josephson currents $I(\phi)$ in unit of $\frac{2e}{\hbar}$ for the magnetic JD for $M=0.2$ and $M=0.4$ and the inset shows the setup for calculating the current. The other parameters are set as $t_R=t_L=1.0$, $\Delta_1=\Delta_2=0.2$, $t_{RB}=0.8$, $t_{LB}=0.6$, $\alpha=0.2$, $\mu=-3.0$. In the calculation, we set the thickness of SC on both sides to be 20.
			\label{fig3}}
	\end{center}
	\vskip-0.5cm
\end{figure}

Just as above, we first investigate the proximity process. It is widely known that magnetism disfavors spin-singlet pairing and the effect of the magnetism on the transport properties of the JJ was studied in detail before~\cite{rmp_jj}. Hence, a larger $M$ should weaken the proximity effect. 
By comparing the FSs at different $M$ in Fig.\ref{fig3}(a),(b), the spin split FSs at $M=0.2$ are slightly different from that at $M=0.4$ with an even larger $\delta k_F$.
From Fig.\ref{fig3}(c), the effective singlet pairing strength along the TB FS for $M=0.2$ is larger than the case with $M=0.4$. Assuming $M_0$=0.3 and with a proper $c_{M}$ and magnetic field strength $B_0$, we can have $M(-B_0)=0.2$ and $M(B_0)=0.4$.
Hence, if the external magnetic field could change the value of $M$, the proximity region can be adjusted.
The Josephson current of the above JD can be also calculated by Eq.~\ref{current}. From Fig.\ref{fig3}(d), the critical current $I_c$ for  $M=0.2$  is larger than the case with  $M=0.4$.  Therefore, an $I_{c}(B)\neq I_{c}(-B)$ phenomenon can be achieved by the external magnetic field and the magnetic order inside the TB layer.
Here, in the calculation for both Fig.~\ref{fig2} and Fig.~\ref{fig3}, we fix the chemical potential $\mu=-3.0$ and the existence of the nonreciprocal effect does not depend on the choice of $\mu$, which only modifies the size of the Fermi surface. This change of the Fermi surface only causes the critical current for both directions to increase or decrease at the same time, which does not qualitatively affect the nonreciprocal effect~\cite{sm}.

\begin{figure}
	\begin{center}
		\fig{3.4in}{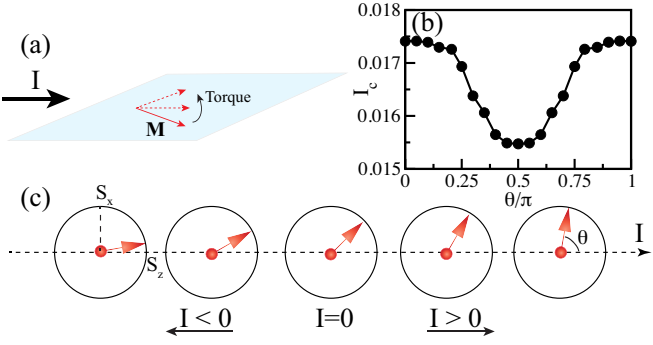}\caption{(a) The TB layer contains magnetic order $M$. The current $I$ induces the magnetic direction rotation through a spin-orbit torque. (b) The critical current $I_c$ in unit of $\frac{2e}{\hbar}$ as a function of $\theta$ the angle of the internal magnetization with respect to the z-axis. This $I_c$ anisotropy is owing to SOC. Here, the magnitude of the magnetization is fixed to $M_0$=0.2 so that $M_z=M_0\cos{\theta}$ and $M_x=M_0\sin{\theta}$. The other parameters are set as $t_R=t_L=1.0$, $\Delta_1=\Delta_2=0.2$, $t_{RB}=t_{LB}=1.0$, $\alpha=0.2$, $\mu=-3.0$. In the calculation, we set the thickness of SC on both sides to be 20 and the thickness of the TB to be 5.
(c) A schematic plot of the current induced JD process. In positive direction, $\theta$ increases towards $S_x$ and in negative direction, $\theta$ decreases towards $S_z$. 
			\label{torque}}
	\end{center}
	\vskip-0.5cm
\end{figure}

The results shown in Fig.~\ref{fig3} demonstrates that a nonreciprocal behavior of the Josephson current could be realized if the internal magnetization can be modified through some time-reversal breaking external field with opposite directions.
Besides the magnetic field induced nonreciprocal effect, the current flow is another efficient way towards realizing  the ${\cal T}$ breaking JD. Especially, the most striking  phenomenon in NSB heterostructure is the nonreciprocal transport depending on the current direction \cite{ali}. 
Since current breaks ${\cal T}$ symmetry, the TB layer must break ${\cal T}$ symmetry for a current controlled JD.
The current flow can be used to tune the magnetic order, which is widely used in spintronics \cite{spintronics1,spintronics2,rashba_review,sot_review}. This idea can be applied to $\cal T$ breaking JD. For example, the combination of  current $iJ f_ic^{\dagger}_j$ and a Kane-Mele type SOC term $i \lambda c_j\sigma_z f_i^\dagger$ \cite{kane-mele} can induce an effective spin order term $-J\lambda f_i<c^{\dagger}_j c_j>\sigma_z f_i^\dagger$ to compete with internal magnetic order.

Besides the magnetic amplitude, another important property used in spintronics is its spin direction $\hat{\bf{n}}$ through spin-orbit torque (SOT) effect~\cite{sot_review}. 
SOT reverses the magnetization direction by electrical current flow owning to spin-orbit coupling, as illustrated in Fig.\ref{torque}. This spin manipulating mechanism has been widely discussed and observed in bulk ferromagnets, antiferromagnets and multilayer heterostructures etc.~\cite{Manchon08,Manchon09,afm,afm2,MacDonald,SHE,Chernyshov,Mihai10,Mihai11}.
We consider the barrier contains internal magnetic order $M$ lying in the x-z plane, whose spin direction is labeled as $\theta$ relative to the $s_z$ axis shown in Fig.\ref{torque}(c).
To simplify our discussion, we further assume that the direction $\theta$ couple to the current by a phenomenological linear equation $\theta=\theta_0+\beta_{\theta} I$, where the initial $\theta_0$ equals to $\frac{\pi}{4}$ and $\beta_{\theta}>0$ is the effective coupling. Hence, if $I>0$, $\theta$ increases towards the $s_x$ axis. If $I<0$, $\theta$ decreases towards the $s_z$ axis, as illustrated in Fig.\ref{torque}(c).
On the other hand, owing to SOC, spin rotation symmetry is broken with magnetic anisotropy. Then, the critical current $I_c$ of JJ becomes $\theta$ dependent. As shown in Fig.\ref{torque}(b), $I_c$ is minimal when $\textbf{M}$ lies in the x direction ($\theta=\frac{\pi}{2}$) and when $\textbf{M}$ is along the z direction ($\theta$=0), $I_c$ reaches the maximum.
From the current manipulation of $\theta$ in Fig.\ref{torque}(c), the critical current in the negative direction $I_{c-}$ is large than that in the positive direction $I_{c+}$, which shows the ${\cal T}$ breaking diode effect using the current flow.

\begin{figure*}
	\begin{center}
		\fig{7.0in}{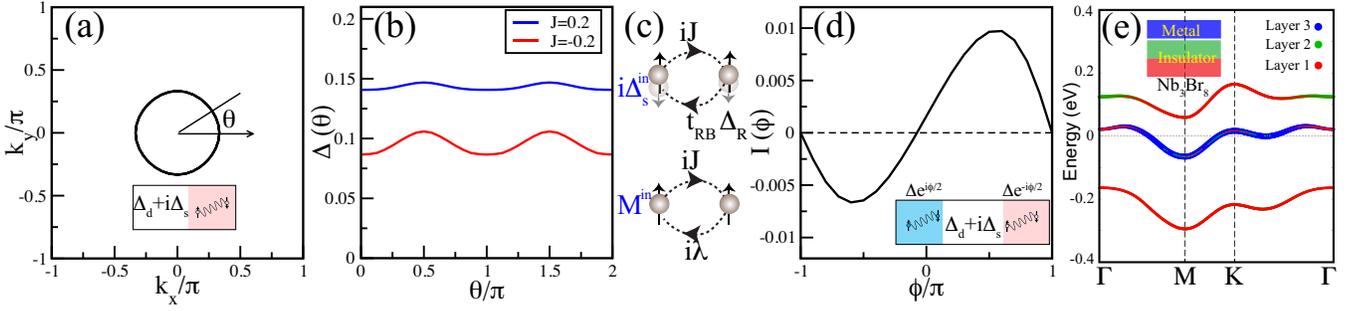}\caption{(a) Fermi surface for TB with $\Delta_d+i\Delta_s$ pairing. The inset illustrates the proximity process leading to the pairing amplitude in (b).  (b) Spin singlet pairing strengths along each FS for $J=\pm0.2$. $\theta$ is the angle along each FS as defined in (a). (c) Perturbation process for pairing and magnetism respectively. In the top panel, a current term $iJ f_ic^{\dagger}_j$, a pairing term $c^{\dagger}_{j\sigma} c^{\dagger}_{j\bar\sigma}$ followed by $ t_{RB} c^{\dagger}_jf_i$ induces $i\Delta_{s}^{in}$ pairing. In the bottom panel, a current term $iJ f_ic^{\dagger}_j$ following SOC $i \lambda c_j\sigma_z f_i^\dagger$ induces spin polarized $M^{in}\sigma_z$.
			(d) The Josephson currents $I(\phi)$ in unit of $\frac{2e}{\hbar}$ for the $\Delta_d+i\Delta_s$ JD. The other parameters are set as $\Delta_{s0}=0.05$, $\Delta_d=0.03 $, $t_R=t_L=1.0$, $\Delta_R=\Delta_L=0.2$, $t_{RB}=0.8$, $t_{LB}=0.4$, $\mu=-3.0$. In the calculation, we set the thickness of SC on both sides to be 20. (e) Band structures for Nb$_3$Br$_8$ TB with wavefunction projections onto each layer. From wavefunction projection, layer 1 and layer 2 strongly couple with each other and form an insulator with band gap. Layer 3 is metallic with bandwidth 0.1eV. 
			\label{fig4}}
	\end{center}
	\vskip-0.5cm
\end{figure*}

Beyond the internal magnetism of the TB layer, the TB layer can  host  ${\cal T}$  breaking  superconducting  at low temperatures. 
Normally,  ${\cal T}$  breaking  SC can happen 
as an instability driven by  multiple order competition and correlation, like the $d+id$ SC in $1/4$ doped graphene \cite{chubukov,qhwang12} and the $d+is$ SC when a d-wave SC coexists with an s-wave SC \cite{ting,shiba,joynt}.
For simplicity, we take a $d+is$ wave SC in the TB layer as an example by assuming TB favors a d-wave SC owing to correlation. More general cases for pure $is$, $s+ip$ etc. are discussed in the supplemental materials \cite{sm}.

Following the above procedure, the proximity process for TB with a $d_{x^2-y^2}+is$ wave pairing ($ i \Delta_{s} \sum_i f_{i\uparrow}f_{i\downarrow}+\Delta_{d}\sum_{<ij>}(-1)^{i_y-j_y}c_ic_j+h.c.$) and $\mathbf{\Delta}_{2}$ under current flow $iJ\sum_{<i,j>}c^{\dagger}_if_j+h.c.$ is calculated. Importantly, the current $J$  term can induce an $i \Delta_s^{in}$ component towards the TB layer through the proximity process as well.
This effect can be understood from a perturbation approach, as illustrated in the up panel of Fig.\ref{fig4}(c). $iJ f_{i\sigma}<c^{\dagger}_{j\sigma}c^{\dagger}_{j\bar\sigma}> t_{RB} f_{i\bar\sigma}$ perturbation process induces an effective  $i \Delta_s$ pairing proportional  to $t_{RB}\Delta_{2}J$ since $<c^{\dagger}_{j\sigma}c^{\dagger}_{j\bar\sigma}>\propto\Delta_{2}$.
Fig.\ref{fig4}(b) plots the effective spin singlet pairing amplitudes along FS in Fig.\ref{fig4}(a)  for $J>0$ and $J<0$, which clearly shows the competition between $i \Delta_{s}$ and the induced $i \Delta_{s}^{in}$.
Therefore, the direction of the current flow can tune the proximity region in a nonreciprocal way through such competition which is described schematically in Fig.~\ref{fig1}(c).
Additionally, the Josephson currents are calculated by Eq.~\ref{current} in Fig.\ref{fig4}(d), which shows a  nonreciprocal critical current of the Josephson junction. Hence, a current-controlled JD can be achieved. 
In addition, in this case, the Josephson current is still finite when the phase bias vanishes, which realizes the anomalous Josephson effect originally related to the $\phi_0$ Josephson junctions~\cite{buzdin_2008}, although the physical origin of the time-reversal breaking comes from the unconventional superconducting pairing which is different from the $\phi_0$ junctions studied before~\cite{buzdin_2008,yuli_2013,yuli_2014,julia_2015,kou_2016,Marco_2019,phi0_2013,phi0_2016,phi0_2017,phi0_2018,phi0_2020,phi0_2021,phi0_2022}.
Moreover, this diode effect disappears if we set $t_{RB}=t_{LB}$, which shows that both inversion symmetry breaking achieved by setting $t_{RB}\neq t_{LB}$ here and ${\cal T}$ symmetry breaking achieved by $d+is$ pairing here are necessary to realize this diode effect.
Because of this symmetry requirement, this nonreciprocal response can also be used to test whether an SC breaks ${\cal T}$ symmetry.

In short, by engineering the ${\cal T}$ broken tunneling barrier via magnetic order and  ${\cal T}$ breaking superconductivity, we demonstrate that the critical current of JJ can be tuned by the current. This mechanism can be used to construct JD with $I_{c+}\neq I_{c-}$. 

\section{Discussion and Summary}
Now, we can apply our general theory to the JD effect found in the NSB heterostructure. Since NbSe$_2$ is a conventional s-wave SC \cite{nbse2,hess}, the unique feature of NSB JD relies on the Nb$_3$Br$_8$ barrier. Nb$_3$Br$_8$ is found to be an obstructed atomic insulator (OAI) \cite{ali,gao_2022,yuanfeng}, which is a generalization of the SSH chain~\cite{Bern_2017,bern_2022}.
In each conventional cell of bulk Nb$_3$Br$_8$ crystal, there are six sub-layers while the NSB only contains three sub-layers breaking the ${\cal I}$ \cite{ali,nb3br8,yuanfeng}.
Using density functional calculations, we further confirm that the physics of Nb$_3$Br$_8$ is dominated by the Nb $d_{3z^2-r^2}$ orbital. Each monolayer  Nb$_3$Br$_8$ is a half-filled single band metal and two Nb$_3$Br$_8$ layers strongly couple with each other opening up a bilayer band gap.
Then stacking Nb$_3$Br$_8$ layers along the z direction form the Nb$_3$Br$_8$ crystal. Interestingly, the interlayer couplings between neighboring layers form strong and weak coupling bonds alternatively in the Nb$_3$Br$_8$ material, which is similar to the 1-D SSH chain or Rice-Mele chain in Fig.~\ref{figr_m}(a).  
Therefore, for Nb$_3$Br$_8$ barrier layer in NSB heterostructure shown in Fig.~\ref{fig4}(e), layer 1 and layer 2 are insulating with a metallic layer 3 floating on top of them, which is the end property of an obstructed atomic insulator \cite{yuanfeng}. Notice that, this TB structure agrees with the geometric setup discussed in Fig.~\ref{fig1}.
More importantly, the bandwidth $W$ of layer 3 or monolayer Nb$_3$Br$_8$ is quite narrow with $W \sim 96$ meV.  On the other hand, a recent experiment shows the correlation strength of Nb is quite strong with Hubbard U around $0.8\sim1.2$ eV~\cite{qian_2022}. Therefore, Nb$_3$Br$_8$ is a strongly correlated system in addition to its atomic obstructed nature.

For the NSB heterostructure, there are two non-reciprocal phenomena related to the finite critical current difference $\Delta I_c$ and the finite returning current difference $\Delta I_r$  \cite{ali}. 
Hence, we can conclude that the $\cal T$ breaking must take place due to the nonvanishing $\Delta I_c$. 
The origin of this $\cal T$ breaking is far from clear. We propose that this effect is from its strongly correlated flatband of the metallic layer.
It is likely that this TB layer may host a magnetic ground state due to the correlation or unconventional superconductivity \cite{sm,nb3br8-mag}. 
Then the current flow tunes the magnetic ground state or superconductivity as we discussed in the ${\cal T}$ breaking JD section.
However, owing to the high complexity of this heterostructure and strong correlation, the detailed feature of NSB is beyond our work, which calls for further theoretical and experimental investigations.
For the $\Delta I_r$, this nonreciprocal feature is coming from its ${\cal I}$ symmetry breaking.
This inversion symmetry breaking in the obstructed atomic insulator is similar to Rice-Mele chain leading to finite electric polarization. Then,
owing to the charge accumulation of the device capacitance, the voltage potential at the interface can change the proximity region through electric polarization giving rise to $\Delta I_r$ \cite{ali,nagaosa_prb}, as we discussed in the ${\cal I}$ breaking JD section.


For the JD effects using the magnetochiral anisotropy and the asymmetric edge states, both of them are ${\cal T}$ breaking JD \cite{hejun,yuan,ando,law}. Taking the Rashba system as an example, the magnetic field $B_y$ along the y direction will induce a finite momentum shift $q_x$ along the x direction because of the Rashba SOC ($k_x \sigma_y -k_y \sigma_x$). Owing to this finite momentum, the current flow along the x direction behaves differently in the  positive and negative directions, which also shows different proximity processes as above.

In summary, 
we study the general theory for Josephson diodes. Based on symmetry analysis,  there are two types of JDs, ${\cal I}$ breaking JD and ${\cal T}$  breaking JD. For ${\cal I}$ breaking JD, the voltage can be used to control the internal potential dependent quantity, like the Rashba SOC or electric polarization, which leads to $I_c(V)\neq I_c(-V)$ and $I_{r+}\neq I_{r-}$. For ${\cal T}$  breaking JD, the current serves as the controlling parameter, which leads to $I_{c+}\neq I_{c-}$. In this case, the tunneling barrier needs to break ${\cal T}$  in addition to ${\cal I}$ breaking, like the internal magnetism or time-reversal breaking pairing. 
All these results provide a comprehensive understanding of JD physics and lead to general principles of JD designs.
We hope our findings could further stimulate the investigation of Josephson diode effects both theoretically and experimentally.


This work is supported by the Ministry of Science and Technology  (Grant
No. 2017YFA0303100), National Science Foundation of China (Grant No. NSFC-11888101, No. NSFC-12174428), and the Strategic Priority Research Program of Chinese Academy of Sciences (Grant
No. XDB28000000). 
Y.Z. is supported in part by NSF China Grant No. 12004383 and No. 12074276 and the Fundamental Research Funds for the Central Universities.

\bibliography{reference}

\clearpage
\onecolumngrid
\begin{center}
\textbf{\large Supplemental Material: General Theory of  Josephson  Diodes}
\end{center}

\setcounter{equation}{0}
\setcounter{figure}{0}
\setcounter{table}{0}
\setcounter{page}{1}
\setcounter{section}{0}
\makeatletter
\renewcommand{\theequation}{S\arabic{equation}}
\renewcommand{\thefigure}{S\arabic{figure}}
\renewcommand{\bibnumfmt}[1]{[S#1]}
\renewcommand{\citenumfont}[1]{S#1}

\twocolumngrid

\section{General Consideration}
\begin{figure}[htb]
	\centerline{\includegraphics[width=0.5\textwidth]{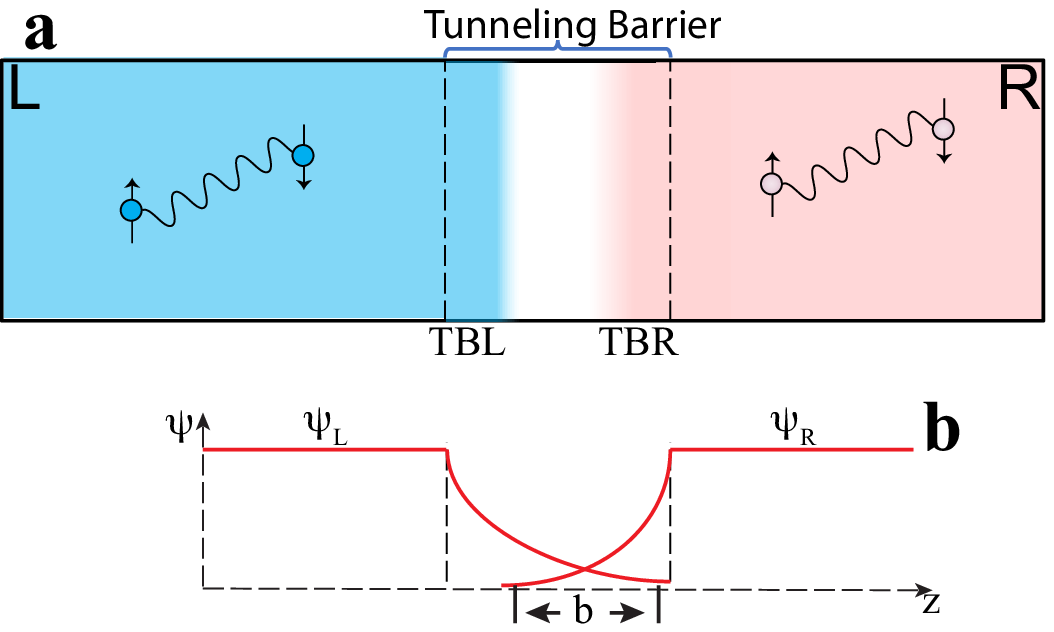}} \caption{ \textbf{a} The general geometric setup for a Josephson diode with two SCs sandwiched with tunneling barrier. Here, the inversion breaking generally gives two proximity regions labeled as TBL and TBR. If the TBL and TBR can be controlled, a Josephson diode can be achieved.
		\textbf{b} The Ginzburg-Landau (GL) description for Josephson junction. The length b is a phenomenological length describing the overlap between GL field $\psi_L$ and $\psi_R$. 
		\label{figs1} }
\end{figure}

Generally speaking, the general setup for a Josephson diode (JD) does not need to be formed by an insulator and a metal in the extreme limit discussed in the main text. A general JD is inversion broken with two proximity regions TBL and TBR, as shown in Fig.\ref{figs1}\textbf{a}. If the external field controls the TBL and TBR asymmetrically, a nonreciprocal transport can be achieved. From the Ginzburg-Landau point of view, the Josephson junction is determined by
the right and left GL fields $\psi_L$, $\psi_R$ and their effective coupling length $b$, as shown in Fig.\ref{figs1}\textbf{b}. If the $b$ is controlled by external field, a JD is formed.

\section{Crystal structures and electronic structures of N$b_3$B$r_8$}

\begin{figure}[htb]
	\centerline{\includegraphics[width=0.5\textwidth]{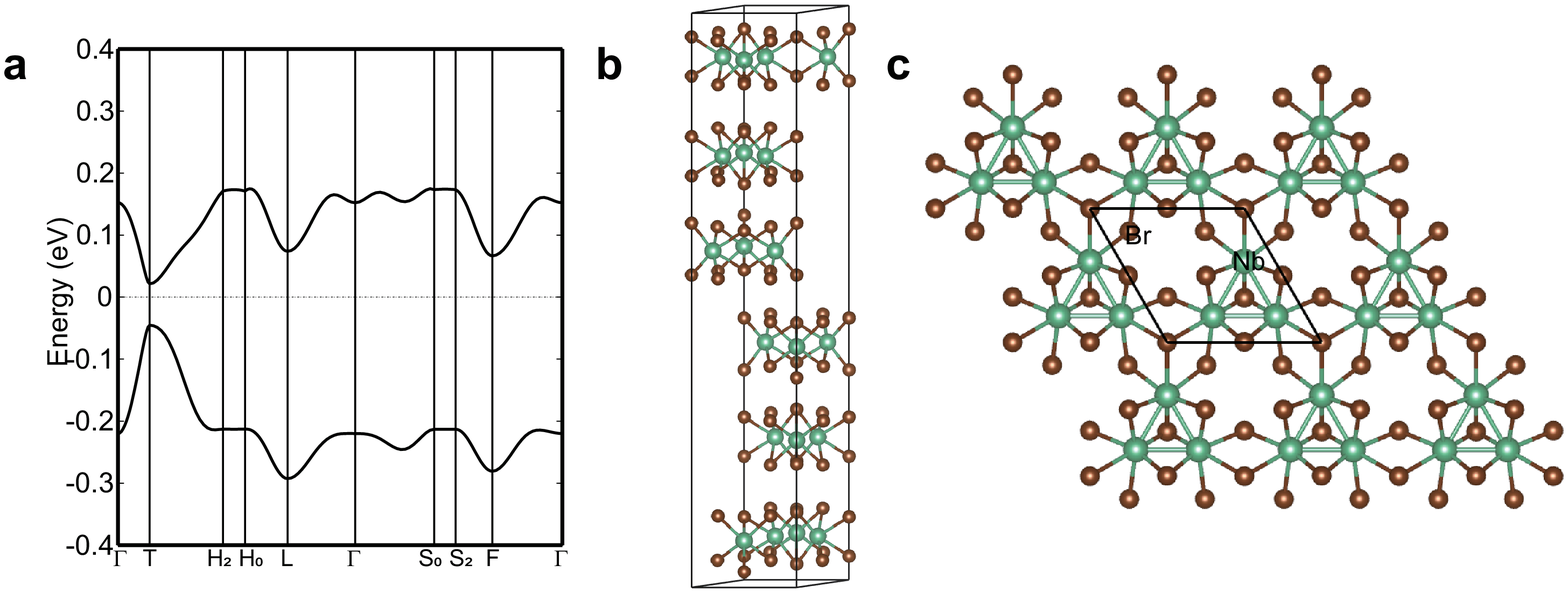}} \caption{(color online) \textbf{a}, The DFT-calculated band structure of \ce{Nb3Br8}. \textbf{b}, The crystal structure of \ce{Nb3Br8} from side view. \textbf{c}, The structure of one layer in \ce{Nb3Br8}.
 \label{bulk} } 
\end{figure}

The crystal structure of \ce{Nb3Br8} is shown in Fig.\ref{bulk}\textbf{b}: six $\ce{Nb3Br8}$ layers stack along [001] direction, forming its conventional cell. In $\ce{Nb3Br8}$, Nb is coordinated with 6 Br atoms and \ce{NbBr6} octahedral complexes connect each other by sharing edges. Thus, the \ce{Nb3Br8} monolayer can be viewed as a distorted version of \ce{CdI2}-type layer with $\frac{1}{4}$ vacancies. Moreover, Nb atoms in \ce{Nb3Br8} will trimerize, as shown in Fig.\ref{bulk}\textbf{c}, leading that half of Nb-Nb bonds shorten (2.87 \AA) while the other half of bonds elongate (4.23 \AA). As a result, Nb atoms form a distorted kagome lattice in \ce{Nb3Br8} monolayer. Although there are six layers in \ce{Nb3Br8} conventional cell, there are only two inequivalent layers in its primitive cell because of its $12R$ stacking structure\cite{habermehl2010triniobiumoctabromide}. This is consistent with the result of the band calculation: there are two bands around the Fermi level, attributed to two distinctive Wannier functions (WFs) centered in $Nb_3$ triangles in two layers. These two bands are relatively flat, owing to the relatively long Nb-Nb bonds between triangles (4.23 \AA).

\begin{figure}[htb]
	\centerline{\includegraphics[width=0.5\textwidth]{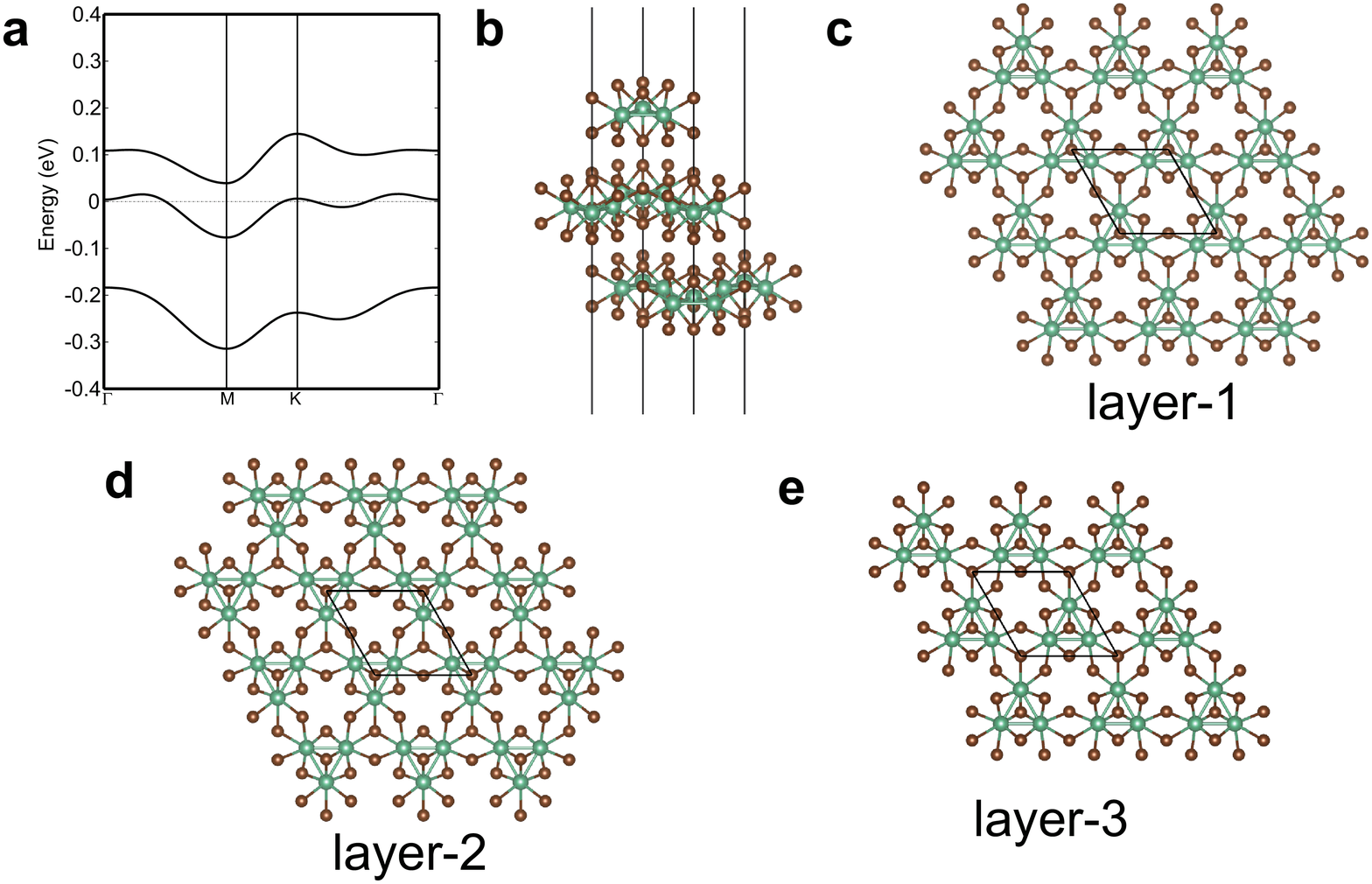}} \caption{(color online) \textbf{a}, The DFT-calculated band structure of trilayer \ce{Nb3Br8}. \textbf{b}, The crystal structure of trilayer \ce{Nb3Br8} from side view. \textbf{c}-\textbf{e}, The crystal structure of each layer in trilayer \ce{Nb3Br8}.
		\label{tri} }
\end{figure}

We build a trilayer \ce{Nb3Br8} model with 40 \AA$ $ thick vacuum layer, which has been realized experimentally\cite{wu2021realization}. The scenario changes here. The WF in layer-1 and the WF in layer-2 are centered in the same site, as shown in Fig.\ref{tri}\textbf{c},\textbf{d}. These two WFs strongly coupled, forming the highest and lowest band in Fig.\ref{tri}\textbf{a}. The rest layer is metallic, forming the half-occupied band in Fig.\ref{tri}\textbf{a}.

\section{Computational details of the first principle calculation}
Our density functional theory (DFT) calculation is performed for the triplelayer Nb$_3$Br$_8$  together with built-in 40 \AA$ $ thick vacuum layer by Vienna ab initio simulation package (VASP) code\cite{kresse1996} with the projector augmented wave (PAW) method\cite{Joubert1999}. The Perdew-Burke-Ernzerhof (PBE)\cite{perdew1996} exchange-correlation functional is used in our calculation. The kinetic energy cutoff is set to be 600 eV for expanding the wave functions into a plane-wave basis. The energy convergence criterion is $10^{-7}$ eV and the $\Gamma$-centered \textbf{k}-mesh is $12\times12\times2$. The triplelayer Nb$_3$Br$_8$ is fully relaxed while forces are minimized to less than 0.001 eV/{\AA}. The zero-damping DFT-D3 method is used to consider vdW interaction\cite{grimme2010consistent}. 

Wannier90 code\cite{mostofi2008wannier90,Marzari2012} is employed to calculate maximally localized Wannier functions (MLWFs) centered in Nb$_3$ triangular clusters.

\section{Details of the tight-binding model calculation}
In our model calculation, we consider a cubic lattice with Josephson current direction along $z$ axis, so that the system is still translation invariant in the $x$-$y$ plane. We use a 20$\times$20 k-mesh in the $x$-$y$ plane and consider 41 unit cells along $z$ direction which consists of right and left superconducting regions with the thickness of 20 and one single layer representing the tunneling barrier (TB). We have performed the calculation with denser k-mesh and larger thickness along the $z$ direction and the results do not change qualitatively.
In the calculation for Fig.6 of the main text, we increase the thickness of TB to 5 to emphasize the effect of the magnetization M in TB.

\subsection{Chemical potential effect}
\begin{figure}[htb]
	\centerline{\includegraphics[width=0.5\textwidth]{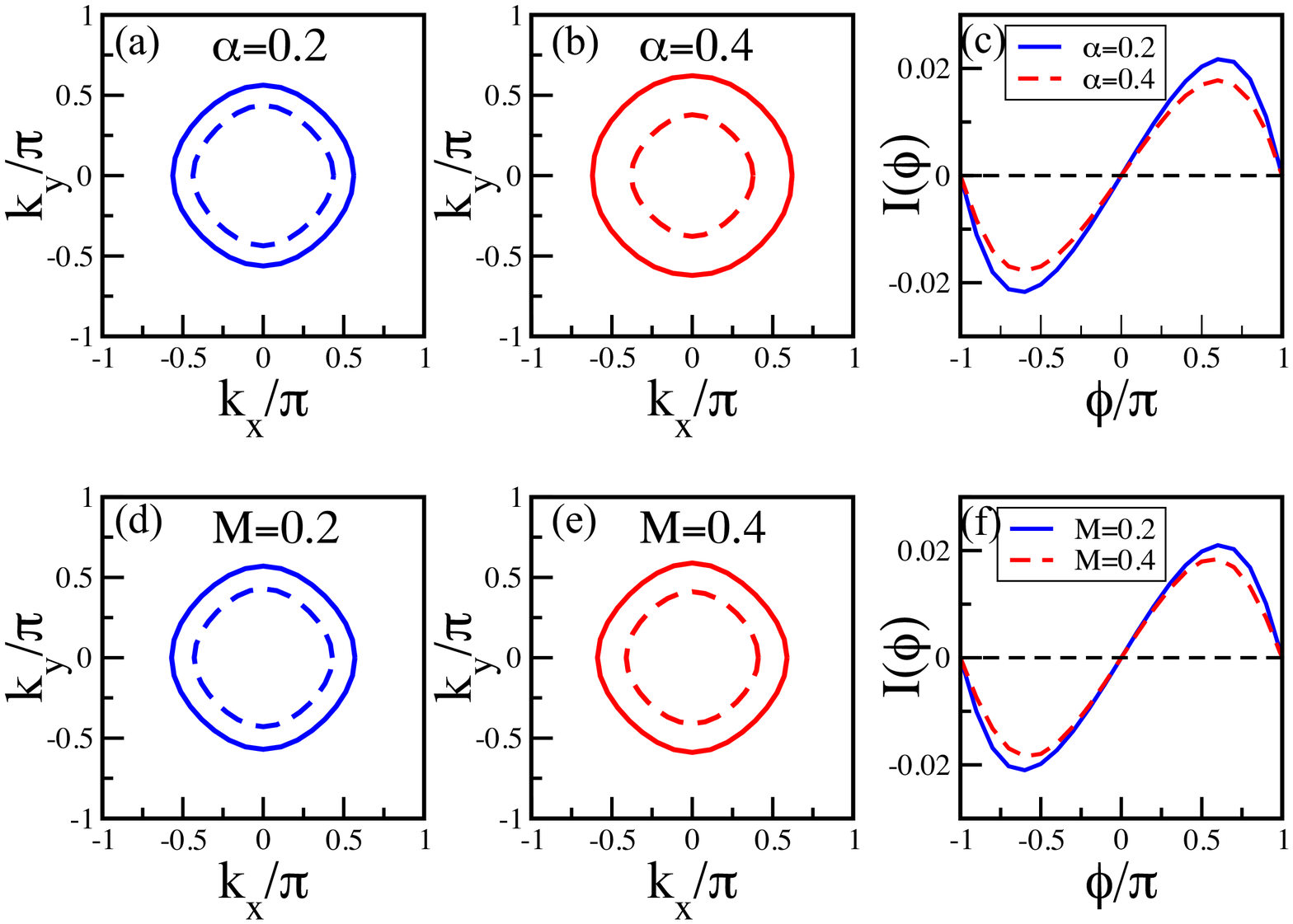}} \caption{(color online) Results of similar calculations as Fig.3 and Fig.5 in the main text except for a different chemical potential $\mu$=-2.0 showing qualitatively the same behavior of the diode effect.  \textbf{(a-c)} Fermi surface of TB with $\alpha$=0.2,0.4 and the corresponding current phase relation for the Josephson junction as compared with Fig.3(a,b,d) in the main text. \textbf{(b-d)} Fermi surface of TB with M=0.2,0.4 and the corresponding current phase relation for the Josephson junction as compared with Fig.5(a,b,d) in the main text.
		\label{figs5}}
\end{figure}

In the calculation showing the diode effects in Fig.3 and Fig.5 of the main text, the chemical potential is fixed at $\mu=-3.0$. In order to show the effect of $\mu$, we perform the same calculation with a different chemical potential as $\mu=-2.0$ and as shown in Fig.~\ref{figs5}, we can see that for $\mu=-2.0$, the critical currents for both positive and negative direction increase, due to the larger size of the Fermi surface, which does not change the diode effect qualitatively.

\subsection{More general $\cal{T}$ breaking case}
\subsubsection{Pure i$s$ wave pairing case}
In the main text, we show the Josephson diode effect of TB with a $d_{x^2-y^2}+is$ wave pairing. Here we show in Fig.\ref{figs3}  that a similar effect can also be achieved in the case of TB with pure i$s$ wave pairing since the key physics is the competition between the induced $i\Delta_s^{in}$ and $i\Delta_s$ in the TB. In this calculation, the parameters used are the same as those used in Fig.4 of the main text except that the d-wave pairing strength $\Delta_d$ is set to 0 to realize a pure  i$s$ wave pairing in TB.

\subsubsection{$s$+ i$p$ wave pairing case}
\begin{figure}[htb]
	\centerline{\includegraphics[width=0.5\textwidth]{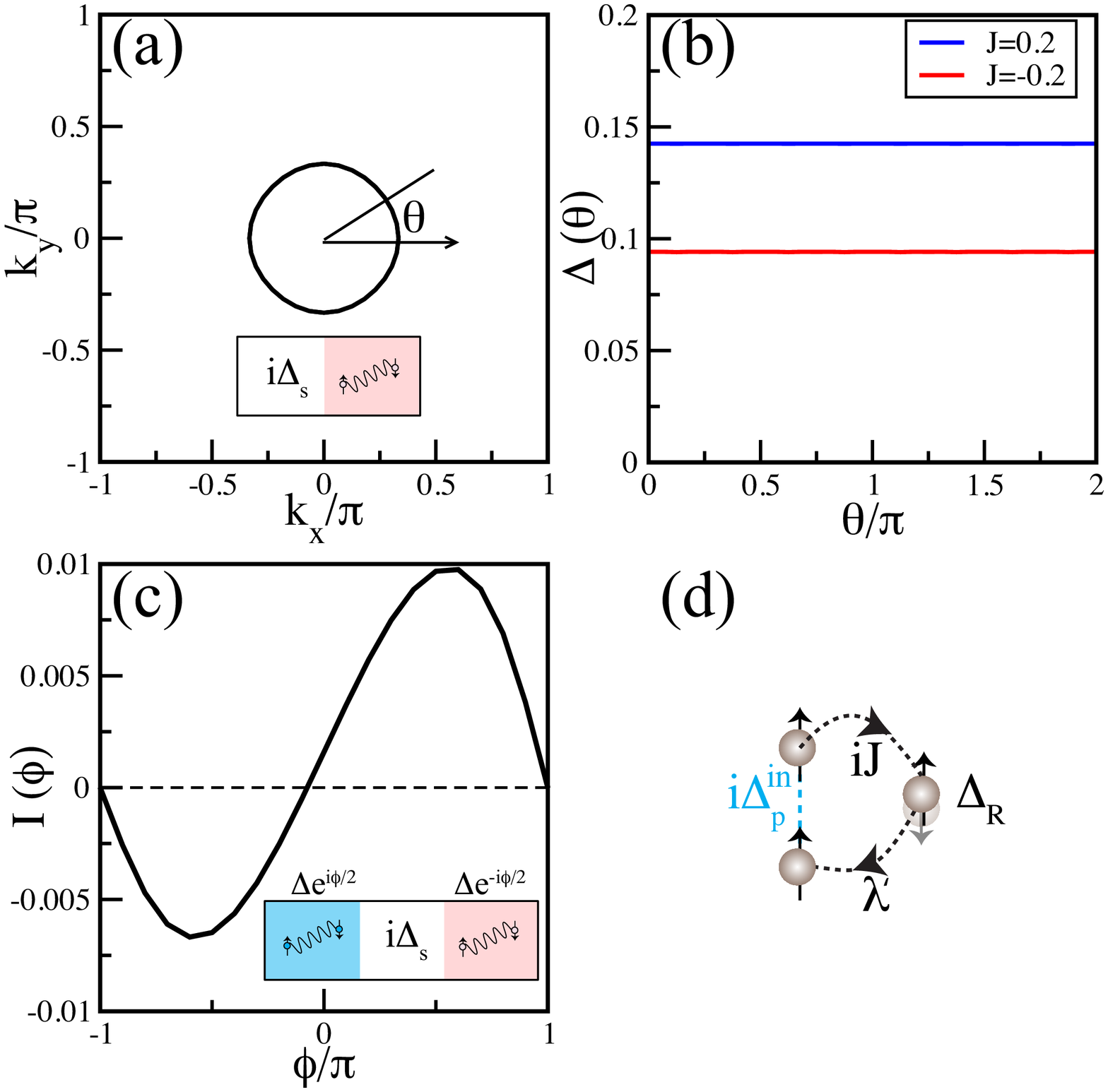}} \caption{(color online) \textbf{a}, Fermi surface for TB with $i\Delta_s$ pairing. The inset illustrates the proximity process leading to the pairing amplitude in \textbf{b}.  \textbf{b}, Spin singlet pairing strengths along each FS for $J=\pm0.2$. $\theta$ is the angle along each FS as defined in \textbf{a}. \textbf{c}, The Josephson currents $I(\phi)$ in unit of $\frac{2e}{\hbar}$ for the $i\Delta_s$ JD. \textbf{d}, Perturbation process for $i\Delta_p$  pairing. A current term $iJ f_{i\sigma}c^{\dagger}_{i'\bar{\sigma}}$, a pairing term $c^{\dagger}_{i'\bar{\sigma}} c^{\dagger}_{i'\bar\sigma}$ followed by $ \lambda' c^{\dagger}_{i'\bar{\sigma}}f_{j\bar{\sigma}}$ SOC induces $i\Delta_{p}^{in}$ pairing. In this calculation,  the parameters used are the same as those used in Fig.4 of the main text except that the d-wave pairing strength $\Delta_d$ is set to 0
		\label{figs3} }
\end{figure}

If the TB has $s+ip$ wave pairing, the system can also have a diode effect through the $p$ wave pairing induced by the current. The mechanism of the induced $p$ wave pairing can be understood from the following perturbation process: $iJf_{i\sigma}<c^{\dagger}_{i^{\prime}\sigma} c^{\dagger}_{i^{\prime}\bar\sigma}> \lambda^{\prime} f_{j\sigma}$ where we have introduced a nearest neighbour spin-orbit coupling $\lambda^{\prime}$ between the fermion $f_{j\sigma}$ in the TB and the fermion $c^{\dagger}_{i^{\prime}\bar\sigma}$ in the bulk superconductor. This induced $p$ wave pairing amplitude $\Delta_p^{in}$ is proportional to $iJ\lambda^{\prime}\Delta_R$ since $<c^{\dagger}_{j\sigma}c^{\dagger}_{j\bar\sigma}>\propto\Delta_{R}$, which can compete with the $s$+ i$p$ wave pairing in the TB depending on the sign of the current. This process is illustrated schematically in Fig.\ref{figs3}\textbf{d}.

Another contribution of the effective $p$ wave pairing comes from the even higher order perturbation process that involves the Rashba coupling inside the TB, which comes from the term $iJf_{i\sigma}c^{\dagger}_{i^{\prime}\sigma}tc^{\dagger}_{i^{\prime}\bar\sigma}f_{i\bar\sigma}(-i\alpha f^{\dagger}_{i\bar\sigma}(\pmb{\sigma}\times\mathbf{d}_{ij})_z^{\bar\sigma\sigma}f_{j\sigma})\propto Jt\alpha\Delta_R<f_{i\bar\sigma}f^{\dagger}_{i\bar\sigma}>(\pmb{\sigma}\times\mathbf{d}_{ij})_z^{\bar\sigma\sigma}f_{i\sigma}f_{j\sigma}$, so that the effective $p$ wave pairing amplitude is proportional to $Jt\alpha\Delta_R$ and has the $p_x+ip_y$ like phase structure. 

\subsection{Rice-Mele chain}
The polarization of the Rice-Mele chain consists of the contributions of both ions and electrons. For the ions, we can treat them as point charges and use the classical method to describe the polarization. For the electrons, since the Wannier functions are localized,  we can treat them as being located at the average position of the Wannier function, called the Wainner center. The polarization of electrons can be defined using the Wannier center associated with the Berry phase of the occupied band and is expressed as \cite{Asbth2016} 
\begin{equation}
    p_{e}=\frac{i}{2 \pi} \int_{-\pi}^{\pi} d k \braket {u(k)} {\partial_{k} u(k)} \text {,}
\end{equation}
in which $u(k)$ is the wave function of the occupied band in Rice-Mele model.\par
Then the polarization of the chain is expressed as a sum over the contributions of point charged ions and the electrons at the  Wannier center of the occupied band \cite{SPALDIN20122}
\begin{equation}
    p(\delta t) = \frac{1}{a}\left(\sum_{i}\left(q_{i} x_{i}\right)^{i o n s} - 2 p_e\right) \text{,}
\end{equation}
where $a$ is the length of the unit cell and the coefficient 2 means there are two electrons in a unit cell. $q_i$ and $x_i$ are the charge and position of ions in a unit cell. The $q_i$ are both +e because the negative charges are described by the polarization of electrons. \par
Since $p$ at $\delta t=0$ is zero with inversion symmetry, we can use it as a reference point. Then the polarization (in unit of e) of the chain is
\begin{equation}
    p = p(\delta t) - p(0) = \frac{1}{a}\left(\sum_{i}\left(\delta x_{i}\right)^{i o n s} - 2 \left(p_e(\delta t)-p_e(0) \right) \right) \text{,}
\end{equation}
where $\delta x_i$ is the deviation from the neutral position of the ions.

\end{document}